\documentclass[prd,nofootinbib,amsfonts,notitlepage]{revtex4-1}
\usepackage{hyperref}
\usepackage{graphicx,bm,amsmath,color}
\usepackage[latin1]{inputenc}
\usepackage{bbold}
\usepackage{wasysym}
\usepackage[T1]{fontenc}
\newcommand{\be}{\begin{equation}}
\newcommand{\ee}{\end{equation}}
\newcommand{\beq}{\begin{eqnarray}}
\newcommand{\eeq}{\end{eqnarray}}

\newcommand{\into}{\leftrightarrow}
\newcommand{\qq}{\quad \quad}

\begin{document}

\title{Modified energy-momentum conservation laws and vacuum Cherenkov radiation}

\author{J.M. Carmona}
\affiliation{Departamento de F\'{\i}sica Te\'orica,
Universidad de Zaragoza, Zaragoza 50009, Spain}
\author{J.L. Cort\'es}
\affiliation{Departamento de F\'{\i}sica Te\'orica,
Universidad de Zaragoza, Zaragoza 50009, Spain}
\author{B. Romeo}
\email{jcarmona@unizar.es, cortes@unizar.es, bromeo@unizar.es}
\affiliation{Departamento de F\'{\i}sica Te\'orica,
Universidad de Zaragoza, Zaragoza 50009, Spain}

\begin{abstract}
We present a general parametrization for the leading order terms in a momentum power expansion of a non-universal Lorentz-violating, but rotational invariant, kinematics and its implications for two-body decay thresholds. The considered framework includes not only modified dispersion relations for particles, but also modified energy-momentum conservation laws, something which goes beyond effective field theory. As a particular and relevant example, bounds on the departures from special relativistic kinematics from the non-observation of vacuum Cherenkov radiation are discussed and compared with those obtained within the effective field theory scenario.

%The results for two-body decay thresholds due to a non-universal Lorentz-violating, but rotational invariant, kinematics are presented.
%The ingredients include both modified dispersion relations for particles and modified energy-momentum conservation laws, something which goes beyond the effective field theory framework. Bounds on the departures from special relativistic kinematics from the non-observation of vacuum Cherenkov radiation are discussed. 
%The results for two-body decay thresholds due to a modified kinematics including the possibility of a non-universal kinematics with modified dispersion relations for the particles and modified energy-momentum composition laws are presented. Bounds on the departures from special relativistic kinematics from the non-observations of vacuum Cerenkov radiation are discussed. 
\end{abstract}

\maketitle

\section{Introduction}

% Motivation: LIV, umbrales, vacuum Cherenkov radiation de interes fenomenologico

Over the last years there has been a strong interest in experimental tests of Lorentz invariance. A violation of this symmetry is suggested by quantum gravity ideas and models, although there exist other theoretical motivations for it~\cite{Mattingly:2005re}. From a quantum gravity perspective, the natural energy scale which controls this possible symmetry violation is the Planck mass, $M_P$, which unfortunately is a very large scale compared to present attainable energies. However, this does not a priori preclude the observation of some of its effects at lower energies. Quantum gravity phenomenology   
\cite{Liberati2011,AmelinoCamelia:2008qg} is a rather new field which tries to identify the phenomenological windows where this could happen. The main idea that helps to understand how effects of quantum gravity could possibly have some phenomenological relevance is that tiny effects can be amplified in specific contexts. This is in fact what happens with thresholds in reactions: they are controlled by masses, even in the ultra-relativistic limit, where they are very small with respect to the relevant energies. Tiny quantum gravity corrections could then modify existent thresholds in the Lorentz invariant theory, produce them in forbidden processes, or eliminate them at large enough (but not Planckian) energies~\cite{Aloisio:2000cm}.

A prototypical example is vacuum Cherenkov radiation $e\to e\gamma$, which is a process forbidden by energy-momentum conservation in special relativity (SR).
This process has been analyzed in Lorentz violating scenarios, where a modified dispersion relation (MDR) for one or both of the two particles, together with the hypothesis of usual energy-momentum conservation, can change this conclusion~\cite{Jacobson:2002hd,Jacobson:2005bg,Saveliev:2011vw,Anselmi:2011ae}. Such an hypothesis is natural in the effective field theory (EFT) framework (such as the standard model extension of Ref.~\cite{Colladay:1998fq}), which is the context normally assumed in this type of studies. One of the reasons for this assumption is that EFT provides a specific realization of local spacetime translational invariance (at least, above a certain length scale) which justifies the use of standard energy-momentum conservation~\cite{Jacobson:2005bg}. Besides this, EFT is also a dynamical theory where one can make explicit computations of matrix elements and reaction rates.

However, EFT could turn out to be an excessively limited framework to analyze phenomenological effects from physics beyond special relativity. This is particularly manifest in contexts where the Lorentz symmetry is deformed, rather than broken, such as in Doubly Special Relativity (DSR) theories~\cite{Amelino-Camelia2002,Magueijo:2002am,AmelinoCamelia:2010pd}. These are scenarios beyond special relativity, but where a relativity principle, that is, an equivalence between inertial observers, still exists. In this case, the relativity principle imposes a specific relationship between the dispersion relation and the composition of energy-momentum in a system of particles~\cite{Carmona2012}, so that it is no longer possible to have a MDR and a standard energy-momentum conservation law (more specifically, the equivalence between observers requires to consider a modified energy-momentum composition law (MCL)~\cite{Amelino-Camelia2002a,Judes:2002bw}).

In contrast to what happens for generic Lorentz invariance violation, DSR theories cannot produce a threshold for particle decays: this is easily seen from the fact that the threshold value of the energy is not relativistically invariant, that is, the energy of the decaying particle could be above or below the threshold for two different observers, and they would disagree about whether the decay takes places or not~\cite{AmelinoCamelia:2002dx,Heyman:2003hs}.
Nevertheless, one could consider modified conservation laws independently from the existence of a relativity principle. In this case, one is out of the EFT scenario and the phenomenological effects are no longer irrelevant.

In this paper we want to go beyond the limitations of a standard energy-momentum conservation law and make some phenomenological studies using a MDR and a MCL. Because of its potential relevance, we will take the process of vacuum Cherenkov radiation as our benchmark for their kinematic implications, but we will not attempt to make any consideration about dynamics. This is obliged since we no longer have the EFT framework at our disposal, and in fact, our premises (a MDR and a MCL) only refer to kinematics, so that any dynamics would require the assumption of further hypotheses. 

Still, one could ask whether there is a spacetime realization from which new, non-additive, energy-momentum conservation laws based on the proposed MCL could emerge as the result of some (modified) concept of translational invariance. Indeed, this is the case. The relative locality framework~\cite{AmelinoCamelia:2011bm,AmelinoCamelia:2011pe} contains a class of models in which the deformations (both of the dispersion relation and the composition laws) arise from modifications of the geometry of momentum space, such as curvature. These theories are indeed nonlocal (or, more precisely, they do not contain an absolute notion of locality of an event), but they do contain a notion of translational invariance associated with the non-linear conservation laws~\cite{Carmona:2011wc}.

Our starting point will therefore be deformed energy-momentum dispersion relations and composition laws. While we will try to consider a scheme as general as possible, we will need to work under certain hypotheses in order to carry out calculations with manageable expressions. First, since we are only interested in the leading modifications to SR (presumably, the only ones with the potential of being detected in the near future), we will consider only the first order corrections in the MDR and the MCL written as a power expansion in the inverse of the ultraviolet energy scale $\Lambda$ which parameterizes the departures from SR (at this point, we do not necessarily assume that $\Lambda$ is the Planck mass $M_P$, but consider a more general scenario). Second, we will make the important algebraic simplification of postulating rotational invariance (in any case, owing to the compactness of the rotation group, this is in general a much better tested symmetry than Lorentz symmetry, which allows for arbitrarily large boosts). With these two hypotheses, we will only need two adimensional parameters $(\alpha_1,\alpha_2)$ to write the general form of the MDR:

\begin{equation}
C^{(a)}(p)=p_0^2-\vec{p}^2+\frac{\alpha_1^a}{\Lambda}p_0^3+\frac{\alpha_2^a}{\Lambda}p_0\vec{p}^2=m^2\,.
\label{eq:MDR}
\end{equation}
Here $m$ is the mass in the special relativistic limit ($\Lambda\to\infty$), the expansion is analytical in the components of the energy-momentum four-vector $(p_0,\vec{p})$, and we have considered a superscript $(a)$ to denote the fact that this MDR can be different for different type of particles. There is no \emph{a priori} reason why the modification to SR should be universal and in fact, from the point of view of the quantum gravity picture, it is reasonable to think that different particles could interact differently with the spacetime microstructure~\cite{Jacobson:2005bg}, depending, for example, on whether they are elementary or not, or on its composition in terms of more elemental constituents. Therefore we leave the possibility for non-universality open.

Finally, the general form of a composition law compatible with rotational invariance is
\begin{equation}
\left[p\oplus q\right]_0 = p_0 + q_0 + \frac{\beta^{ab}_1}{\Lambda} \, p_0 q_0 + \frac{\beta^{ab}_2}{\Lambda} \, \vec{p}\cdot\vec{q} \quad \quad \quad \quad \left[p \oplus q\right]_i = p_i + q_i + \frac{\gamma^{ab}_1}{\Lambda} \, p_0 q_i + \frac{\gamma^{ab}_2}{\Lambda} \, p_i q_0
+ \frac{\gamma^{ab}_3}{\Lambda} \, \epsilon_{ijk} p_j q_k
\label{eq:MCL}
\end{equation}
where $\epsilon_{ijk}$ is the Levi-Civita symbol, a totally antisymmetric tensor, $(a)$ and $(b)$ are the types of particles whose energy-momenta we are composing (we will use the notation $\beta^a\equiv \beta^{aa}$) and it is implemented the condition
\begin{equation}
p \oplus q|_{q=0} = p \quad \quad  \quad \quad p \oplus q|_{p=0} = q\,,
\label{eq:cl0}
\end{equation}
which reasonably says that the composition of two momenta when one is equal to zero is just the only non-zero momentum. With this condition we avoid that the composition $p\oplus q$ contains terms of the type $p_i^2/\Lambda$, of difficult interpretation if $q=0$.

Equations~\eqref{eq:MDR} and~\eqref{eq:MCL} are not invariant under usual Lorentz transformations. This means that the coefficients $(\alpha_i, \beta_i, \gamma_i)$ will be different in coordinate systems related by these transformations. In Sec.~IV we will consider the phenomenological bounds that vacuum Cherenkov radiation imposes on these coefficients. However, since Lorentz invariance is broken, these bounds will be applicable only on the reference system where the analysis is done (laboratory, or Earth-based, reference system). A simple change of coordinates will allow then to express these bounds in a reference system connected with a Lorentz transformation with the laboratory system. However, one could ask whether it is possible to find a new set of coordinate systems (now related by modified Lorentz transformations) where the explicit form of the modified dispersion relation~\eqref{eq:MDR} and modified composition laws~\eqref{eq:MCL} be invariant (i.e., with the same coefficients). We would say in that case that the theory contains a relativity principle. It turns out that this is possible only in cases where there is a specific relation between the $\alpha_i$ coefficients of the MDR and the ($\beta_i,\gamma_i$) of the MCL. This was proven in Ref.~\cite{Carmona2012} for the case of a universal kinematics, and can in fact be extended to the case of non-universal kinematics~\cite{Carmona:2014aba}. In the following, however, we will not include this restriction and will consider a generic form of the MDR and the MCL so that Lorentz invariance is broken, and not simply deformed.

With these basic ingredients, we proceed now to examine the new kinematics in the most simple case: a two-body decay process.

\section{Two-body decay kinematics beyond special relativity}
\label{sec:kinematics}

\subsection{Preliminary considerations} 

The process under study is $A(k)\to C(p)+D(q)$, where $A$ is a particle of type $(a)$ and its momentum is $k$, $C$ is a particle of type $(c)$ and momentum $p$, and $D$ is a particle of type $(d)$ and momentum $q$, and we ask when this process is allowed with the new kinematics introduced in the last Section. 

Firstly, we need to know the law that replaces the usual energy-momentum conservation of SR, once the usual addition of momenta is changed by the MCL Eq.~\eqref{eq:MCL}. This is explicitly constructed in the relative locality framework mentioned in the Introduction, where the conservation law of SR
\begin{equation}
(-k)+p+q=0
\label{eq:SRconslaw}
\end{equation}
is replaced by
\begin{equation}
\hat{k}\oplus p \oplus q=0,
\label{eq:newconslaw}
\end{equation}
where $\hat{k}$ is the \emph{antipode} of $k$, that is, the four-vector that satisfies $\hat{k}\oplus k=k\oplus\hat{k}=0$.\footnote{We are assuming here that it is not necessary to distinguish a left-handed antipode $\hat{k}_L$, $\hat{k}_L\oplus k=0$ from a right-handed antipode $\hat{k}_R$, $k\oplus \hat{k}_R=0$. This is obviously the case if we make the simplification that the antipode of a momentum of type $(a)$ composes as a momentum of type $(a)$.} Note that Eq.~\eqref{eq:newconslaw} contains the composition of \emph{three} momenta. Nevertheless, under the condition that it has to reduce to the composition of two momenta when the other one is equal to zero (something equivalent to Eq.~\eqref{eq:cl0}), the composition of three momenta is completely determined by the coefficients appearing in the composition of two momenta, and Eq.~\eqref{eq:MCL} is generalized to
\begin{subequations}
\begin{align}
[\hat{k} \oplus p \oplus q]_0 &= \hat{k}_0 + p_0 + q_0 + \frac{\beta_1^{ac}}{\Lambda}\hat{k}_0 p_0+ \frac{\beta_1^{ad}}{\Lambda}\hat{k}_0 q_0 + \frac{\beta_1^{cd}}{\Lambda}p_0 q_0+
\frac{\beta_2^{ac}}{\Lambda}\vec{\hat{k}}\cdot \vec{p} + \frac{\beta_2^{ad}}{\Lambda}\vec{\hat{k}}\cdot \vec{q} + \frac{\beta_2^{cd}}{\Lambda}\vec{p}\cdot \vec{q} 
\\
[\hat{k} \oplus p \oplus q]_i &= \hat{k}_i + p_i + q_i + \frac{\gamma_1^{ac}}{\Lambda}\hat{k}_0 p_i+ \frac{\gamma_1^{ad}}{\Lambda}\hat{k}_0q_i  + \frac{\gamma_1^{cd}}{\Lambda}p_0 q_i+
\frac{\gamma_2^{ac}}{\Lambda}\hat{k}_i p_0+ \frac{\gamma_2^{ad}}{\Lambda}\hat{k}_i q_0 + \frac{\gamma_2^{cd}}{\Lambda}p_i q_0 \nonumber \\
& +
\frac{\gamma_3^{ac}}{\Lambda} \epsilon_{ijl} \hat{k}_j p_l + \frac{\gamma_3^{ad}}{\Lambda} \epsilon_{ijl} \hat{k}_j q_l  + \frac{\gamma_3^{cd}}{\Lambda} \epsilon_{ijl} p_j q_l \,.
\end{align}
\label{eq:3MCL}
\end{subequations}

Secondly, we note that the new conservation law Eq.~\eqref{eq:newconslaw} implies an order in the momenta $(\hat{k},p,q)$. This is not surprising since the MCL Eq.~(\ref{eq:MCL}) was already dependent on the order of the momenta, that is, in general $p\oplus q\neq q \oplus p$. This means that we could equally consider $\hat{k}\oplus q\oplus p=0$ or $p\oplus \hat{k}\oplus q=0$ as conservation laws for this process, and the kinematic conditions obtained from these different \emph{channels} will be different. The natural interpretation of this situation, which is exclusive of a MCL, is that if a certain kinematic configuration of momenta is compatible with one of the channels, then it is an allowed configuration for the reaction to proceed. This obviously complicates the analysis of the kinematics of processes beyond the SR case. In the following we will study in detail one of the channels, $\hat{k}\oplus p\oplus q=0$, and then, at the end of this Section and in the next one, we will elaborate on the analysis taking into account all the different channels.

\subsection{Derivation of the equation of the kinematics of the process}

We define now our strategy to get the kinematic consequences of the MDR and the MCL in the $A(k)\to C(p)+D(q)$ process. As in SR, they can be obtained from a single equation. One way to get it is to select one of the particles and use the conservation law Eq.~\eqref{eq:newconslaw} to express the energy and momentum of the selected particle in terms of the three momenta of the rest of the particles, whose energies are written as a function of their momenta by using the corresponding modified dispersion relations. The equation for the kinematics of the process is the relation between the three-momenta of the rest of the particles which follows from imposing the MDR of the selected particle.

From Eqs.~\eqref{eq:newconslaw} and \eqref{eq:3MCL}, we get~\footnote{We will neglect terms proportional to $1/\Lambda^2$ from now on.}
\begin{subequations}
\begin{align}
\hat{k}_0 & = -(p_0+q_0)+\frac{\beta_1^{ac}}{\Lambda}(p_0+q_0)p_0 + \frac{\beta_1^{ad}}{\Lambda}(p_0+q_0)q_0 - \frac{\beta_1^{cd}}{\Lambda}p_0q_0 + \frac{\beta_2^{ac}}{\Lambda}(\vec{p}+\vec{q})\cdot \vec{p} + \frac{\beta_2^{ad}}{\Lambda}(\vec{p}+\vec{q})\cdot \vec{q} - \frac{\beta_2^{cd}}{\Lambda} \vec{p}\cdot \vec{q} 
\label{eq:interm1} \\
\vec{\hat{k}} & =-(\vec{p}+\vec{q})+\frac{\gamma_1^{ac}}{\Lambda}(p_0+q_0)\vec{p} + \frac{\gamma_1^{ad}}{\Lambda}(p_0+q_0)\vec{q} - 
\frac{\gamma_1^{cd}}{\Lambda} p_0 \vec{q} + \frac{\gamma_2^{ac}}{\Lambda}(\vec{p}+\vec{q}) p_0 + \frac{\gamma_2^{ad}}{\Lambda}(\vec{p}+\vec{q}) q_0 - \frac{\gamma_2^{cd}}{\Lambda}\vec{p}\, q_0 \nonumber \\ & - \frac{(\gamma_3^{ac}-\gamma_3^{ad}+\gamma_3^{cd})}{\Lambda}\,\vec{p}\wedge\vec{q}\,.
\label{eq:interm2}
\end{align}
\end{subequations}

Introducing the variables
\begin{equation}
E_p\equiv \sqrt{\vec{p}^2+m_c^2} \quad , \quad E_q\equiv \sqrt{\vec{q}^2+m_d^2}\quad \,,
\label{eq:SRenergies}
\end{equation}
we note that 
\begin{equation}
p_0=E_p+\mathcal{O}\left(\frac{1}{\Lambda}\right) \quad \quad q_0=E_q+\mathcal{O}\left(\frac{1}{\Lambda}\right) \quad \quad 
\vec{p}\cdot\vec{q}=E_p E_q + \frac{m_c^2+m_d^2-m_a^2}{2}+\mathcal{O}\left(\frac{1}{\Lambda}\right)
\label{eq:identities1}
\end{equation}
\begin{equation}
(\vec{p}+\vec{q})\cdot\vec{q}=(E_p+E_q)E_q-\frac{m_a^2+m_d^2-m_c^2}{2}+\mathcal{O}\left(\frac{1}{\Lambda}\right) \quad \quad 
(\vec{p}+\vec{q})\cdot\vec{p}=(E_p+E_q)E_p-\frac{m_a^2+m_c^2-m_d^2}{2}+\mathcal{O}\left(\frac{1}{\Lambda}\right)
\label{eq:identities2}
\end{equation}

When calculating $\hat{k}_0^2-\vec{\hat{k}}^2$, we can now replace $p_0$, $q_0$, $\vec{p}\cdot\vec{q}$, etc, in the terms involving first-order corrections (proportional to $1/\Lambda$) by their zeroth-order equivalent, given by Eqs.~\eqref{eq:identities1} and~\eqref{eq:identities2}. This cannot be done in the zeroth-order terms, where we will find the product $2 p_0 q_0$ after squaring $(p_0+q_0)$ in Eq.~\eqref{eq:interm1}. To write it as a function of $E_p$ and $E_q$ we take the modified dispersion relation for particle $C$:
\begin{equation}
p_0^2=E_p^2-\frac{\alpha_1^c+\alpha_2^c}{\Lambda} E_p^3 + \frac{\alpha_2^c}{\Lambda} E_p m_c^2 \quad \Rightarrow \quad
p_0=E_p-\frac{\alpha_1^c+\alpha_2^c}{2\Lambda} E_p^2+\frac{\alpha_2^c}{2\Lambda} m_c^2\,,
\label{eq:MDRsqrt}
\end{equation}
and an equivalent expression for particle $D$, so that we get
\begin{equation}
2 p_0 q_0 =2 E_p E_q - \frac{\alpha_1^c+\alpha_2^c}{\Lambda} E_p^2 E_q - \frac{\alpha_1^d + \alpha_2^d}{\Lambda} E_p E_q^2 + 
\frac{\alpha_2^c}{\Lambda} E_q m_c^2 + \frac{\alpha_2^d}{\Lambda} E_p m_d^2\,.
\label{eq:2pq}
\end{equation}
 
Finally, from the MDR for the antipode of the momentum of particle $A$,
\begin{equation}
\hat{k}_0^2-\vec{\hat{k}}^2+\frac{\hat{\alpha}_1^a}{\Lambda}\hat{k}_0^3+\frac{\hat{\alpha}_2^a}{\Lambda}\hat{k}_0\vec{\hat{k}}^2-m_a^2=0\,,
\label{eq:MDRantipode}
\end{equation}
we get the following equation:
\begin{equation}
2 E_p E_q - 2\vec{p}\cdot\vec{q}-m_a^2+m_c^2+m_d^2=\hat{O}_3^{A}\,,
\label{eq:master}
\end{equation}
where all the terms proportional to $1/\Lambda$ are contained in the right-hand side, what we have denoted by $\hat{O}_3^{A}$.
This is a very convenient form to write the equation defining the kinematics of the process, since the left-hand side is exactly what one would equate to zero in the SR case. We will call it the \emph{modified kinematics equation} (MKE).

The expression of $\hat{O}_3^{A}$ is
\begin{align}
\hat{O}_3^{A}&=\frac{E_p+E_q}{\Lambda}\left\{(\alpha_1^c+\alpha_2^c)E_p^2+(\alpha_1^d+\alpha_2^d)E_q^2+
(\hat{\alpha}_1^a+\hat{\alpha}_2^a) (E_p+E_q)^2 + 2 (\beta_1^{ac}+\beta_2^{ac}-\gamma_1^{ac}-\gamma_2^{ac})(E_p+E_q)E_p \right. \nonumber \\ 
& \left. + 2 (\beta_1^{ad}+\beta_2^{ad}-\gamma_1^{ad}-\gamma_2^{ad}) (E_p+E_q)E_q-2 (\beta_1^{cd}+\beta_2^{cd}-\gamma_1^{cd}-\gamma_2^{cd}) E_p E_q \right\} + \mathcal{O}\left(\frac{Em^2}{\Lambda}\right)\,.
\label{eq:masteroperator}
\end{align}

\subsection{Some comments about the MKE}
\label{sec:comments}

We see that Eq.~\eqref{eq:masteroperator} contains different combinations of the $(\alpha,\beta,\gamma)$ coefficients appearing in the MDR and the MCL of the three particles $A$, $C$ and $D$.\footnote{Note that the $\gamma_3$ coefficients are absent.} It also contains the coefficients $\hat{\alpha}_1^a$ and $\hat{\alpha}_2^a$ appearing in the MDR of the antipode of the momentum of particle $A$, $\hat{k}$. It should be stressed that the MDR that satisfies $\hat{k}$ is not the same that the one that satisfies $k$, but the $\hat{\alpha}^a$ coefficients are determined by the $(\alpha^a,\beta^a,\gamma^a)$
coefficients. To see this, let us start with the definition of antipode, $k\oplus \hat{k}=0$, which gives
\begin{equation}
k_0=-\hat{k}_0+\frac{\beta_1^a+\beta_2^a}{\Lambda} \hat{k}_0^2 - \frac{\beta_2^a}{\Lambda} m_a^2 \,, \quad \quad \quad
\vec{k}=-\vec{\hat{k}} + \frac{\gamma_1^a+\gamma_2^a}{\Lambda} \hat{k}_0 \vec{\hat{k}}\,.
\label{eq:antipode}
\end{equation}
We have then
\begin{multline}
k_0^2-\vec{k}^2+\frac{\alpha_1^a}{\Lambda} k_0^3 + \frac{\alpha_2^a}{\Lambda}k_0 \vec{k}^2 = 
k_0^2-\vec{k}^2+\frac{\alpha_1^a+\alpha_2^a}{\Lambda} k_0^3 - \frac{\alpha_2^a}{\Lambda}k_0 m_a^2 \\
= \hat{k}_0^2 - \vec{\hat{k}}^2+\frac{1}{\Lambda}\left[-(\alpha_1^a+\alpha_2^a)-2(\beta_1^a+\beta_2^a-\gamma_1^a-\gamma_2^a)\right]\hat{k}_0^3 - \frac{1}{\Lambda}\left[-\alpha_2^a-2(\beta_2^a-\gamma_1^a-\gamma_2^a)\right]\hat{k}_0 m_a^2\,.
\label{eq:interm3}
\end{multline}
Comparing this with Eq.~\eqref{eq:MDRantipode}, we obtain
\begin{equation}
\hat{\alpha}_1^a=-\alpha_1^a-2\beta_1^a\,, \quad \quad \hat{\alpha}_2^a=-\alpha_2^a-2(\beta_2^a-\gamma_1^a-\gamma_2^a)\,.
\label{eq:coefMDRantipode}
\end{equation}

On the other hand, our MKE Eq.~\eqref{eq:master} is a relationship between the three-momenta of the two out-going particles, $\vec{p}$ and $\vec{q}$ (containing the parameters $m_a$, $m_c$ and $m_d$ and the coefficients of the MDR and MCL for all the particles). The reason why this is so is that in order to derive the MKE we selected the first four-momentum appearing in the conservation law Eq.~\eqref{eq:newconslaw}, $\hat{k}$, and obtained the MKE from the MDR of this four-momentum, Eq.~\eqref{eq:MDRantipode}. Obviously, we could have selected a different four-momentum, such as, for example, the one of particle $C$, $p$, then use the conservation law Eq.~\eqref{eq:newconslaw} to write $(p_0,\vec{p})$ as a function of $\vec{\hat{k}}$ and $\vec{q}$ and then get a different MKE from the MDR for $(p_0,\vec{p})$. What one gets in this case is 
\begin{equation}
2(-E_{\hat k}) E_q - 2 \vec{\hat{k}}\cdot \vec{q} - m_c^2+m_a^2+m_d^2 = O_3^{C}\,,
\label{eq:C-MKE}
\end{equation}
where $E_{\hat k}=\sqrt{\hat{k}^2+m_a^2}$. The left-hand side of this alternative MKE is the same as that of the original MKE under the change $E_p \to -E_{\hat{k}}$, $\vec{p} \to \vec{\hat{k}}$ and $m_a^2 \leftrightarrow m_c^2$, while the new $\mathcal{O}(1/\Lambda)$ correction, $O_3^{C}$ (the superscript denotes that we selected the momentum of particle $C$ in order to derive the equation, and the subindex denotes that it is a three-particle process, involving three different momenta), can be obtained from $\hat{O}_3^{A}$ (here the hat denotes that the selected momentum is in fact the antipode of the momentum of particle $A$) by making the same substitutions than for the left-hand side, plus the following replacements:
\begin{equation}
\alpha_1^c \into \hat{\alpha}_1^a \qq \alpha_2^c \into \hat{\alpha}_2^a \qq \gamma_1^{ac} \into \gamma_2^{ac} \qq \beta_1^{ad}\into \beta_1^{cd} \qq \beta_2^{ad}\into \beta_2^{cd}
\qq \gamma_1^{ad}\into \gamma_1^{cd} \qq \gamma_2^{ad}\into \gamma_2^{cd} \,.
\label{eq:replaceC}
\end{equation}

Analogously, had we selected the momentum of particle $D$, $q$, to derive the MKE, we would have got
\begin{equation}
2(-E_{\hat{k}})E_p -  2 \vec{\hat{k}}\cdot \vec{p} - m_d^2+m_a^2+m_c^2 = O_3^{D}\,,
\label{eq:D-MKE}
\end{equation}
where $O_3^{D}$ is obtained from $\hat{O}_3^{A}$ by the replacements $E_q \to -E_{\hat{k}}$, $\vec{q}\to \vec{\hat{k}}$ and $m_a^2 \leftrightarrow m_d^2$, together with
\begin{equation}
\alpha_1^d \into \hat{\alpha}_1^a \qq \alpha_2^d \into \hat{\alpha}_2^a \qq \gamma_1^{ad} \into \gamma_2^{ad} \qq \beta_1^{ac}\into \beta_1^{cd} \qq \beta_2^{ac}\into \beta_2^{cd}
\qq \gamma_1^{ac}\into \gamma_2^{cd} \qq \gamma_2^{ac}\into \gamma_1^{cd} \,.
\label{eq:replaceD}
\end{equation}

Evidently, the three equations~\eqref{eq:master}, \eqref{eq:C-MKE} and \eqref{eq:D-MKE} contain the same physical information (they were derived from the same conservation law); the choice of the MKE in one form or another just depends on the convenience in a given situation (for example, for a discussion involving only the three-momenta of the out-going particles we would choose the first form).

Let us come back to the issue of the different channels for this process; that is, we consider now a different conservation law for the process such as, for example, $p \oplus \hat{k} \oplus q = 0$. The equivalent of Eqs.~\eqref{eq:3MCL} is 
\begin{subequations}
\begin{align}
[p \oplus \hat{k} \oplus q]_0 &= \hat{k}_0 + p_0 + q_0 + \frac{\beta_1^{ca}}{\Lambda}p_0 \hat{k}_0+ \frac{\beta_1^{ad}}{\Lambda}\hat{k}_0 q_0 + \frac{\beta_1^{cd}}{\Lambda}p_0 q_0+
\frac{\beta_2^{ca}}{\Lambda} \vec{p}\cdot \vec{\hat{k}}  + \frac{\beta_2^{ad}}{\Lambda}\vec{\hat{k}}\cdot \vec{q} + \frac{\beta_2^{cd}}{\Lambda}\vec{p}\cdot \vec{q} 
\\
[p \oplus \hat{k} \oplus q]_i &= \hat{k}_i + p_i + q_i + \frac{\gamma_1^{ca}}{\Lambda}p_0\hat{k}_i + \frac{\gamma_1^{ad}}{\Lambda}\hat{k}_0 q_i  + \frac{\gamma_1^{cd}}{\Lambda}p_0 q_i+ 
\frac{\gamma_2^{ca}}{\Lambda} p_i \hat{k}_0 + \frac{\gamma_2^{ad}}{\Lambda}\hat{k}_i q_0 + \frac{\gamma_2^{cd}}{\Lambda}p_i q_0 \nonumber \\
& +
\frac{\gamma_3^{ca}}{\Lambda} \epsilon_{ijl} p_j\hat{k}_l + \frac{\gamma_3^{ad}}{\Lambda} \epsilon_{ijl} \hat{k}_j q_l  + \frac{\gamma_3^{cd}}{\Lambda} \epsilon_{ijl} p_j q_l \,.
\end{align}
\label{eq:3MCLbis}
\end{subequations}
Therefore, the equation of the kinematics in the $p \oplus \hat{k} \oplus q = 0$ channel is obtained from that of the $\hat{k}\oplus p \oplus q=0$ channel by making the following replacements:
\begin{equation}
\beta_1^{ac}\to\beta_1^{ca} \qq \beta_2^{ac}\to\beta_2^{ca} \qq \gamma_1^{ac}\to \gamma_2^{ca} \qq \gamma_2^{ac}\to \gamma_1^{ca}\,.
\label{eq:replacechannel-1}
\end{equation}
The equation for the $p\oplus q \oplus \hat{k}=0$ channel can be obtained from the one for the $\hat{k}\oplus p \oplus q=0$ channel by making the replacements:
\begin{equation}
\beta_1^{ac}\to\beta_1^{ca} \qq \beta_1^{ad}\to \beta_1^{da} \qq \beta_2^{ac}\to\beta_2^{ca} \qq \beta_2^{ad}\to\beta_2^{da} \qq \gamma_1^{ac}\to \gamma_2^{ca} \qq \gamma_2^{ac}\to \gamma_2^{ca} \qq \gamma_1^{ad}\to \gamma_2^{da} \qq \gamma_2^{ad}\to \gamma_1^{da}\,,
\label{eq:replacechannel-2}
\end{equation}
since there are two transpositions, $a\into c$ and $a \into d$, to go from the $(a,c,d)$ order to $(c,d,a)$. Similarly, one can obtain the MKE for the remaining three channels: $\hat{k} \oplus q\oplus p = 0$, $q \oplus \hat{k} \oplus p = 0$, and $q \oplus p \oplus \hat{k} = 0$. 

To conclude this Section, we have to comment on a further complication. We have considered up to now six different channels which result from the different orders of the four-momenta appearing in the conservation law, $\hat{k}$, $p$ and $q$. However, in order to write this law we had to make the choice of taking antipodes for the in-going momentum and leaving the out-momenta untouched. Obviously we could have done conversely: taking antipodes for the out-going momenta and leaving the in-going momentum untouched. This would mean considering the conservation law $k \oplus \hat{p} \oplus \hat{q} = 0$ instead of $\hat{k} \oplus p \oplus q = 0$, and similarly for the rest of the orders of momenta. This is not redundant (ie, the physical consequences are different) since it was the presence of the antipode of $k$ which produced the apparition of the $\hat{\alpha}_i^a$ coefficients in Eq.~\eqref{eq:masteroperator}, and, therefore (according to Eq.~\eqref{eq:coefMDRantipode}), the presence of $(\beta_i^a,\gamma_i^a)$, the coefficients appearing in the composition of momenta of particles of the type $(a)$, in the MKE, which was, on the other hand, completely independent of the coefficients $(\beta_i^c,\beta_i^d,\gamma_i^c,\gamma_i^d)$. On the contrary, these coefficients will appear in the MKE associated to the conservation law $k \oplus \hat{p} \oplus \hat{q} = 0$: we have to consider it, therefore, as a new independent channel. Together with its reorder of momenta, we will have at the end 12 different channels in total along which the reaction can take place.

A direct computation shows that one of the forms in which one can write the MKE for the $k \oplus \hat{p} \oplus \hat{q} = 0$ channel is
\begin{equation}
2(-E_{\hat p})(-E_{\hat q})-2 \vec{\hat{p}}\cdot\vec{\hat{q}}-m_a^2+m_c^2+m_d^2=O_3^A\,,
\label{eq:antipode-MKE}
\end{equation}
where $O_3^A$ can be obtained from $\hat{O}_3^A$ (Eq.~\eqref{eq:masteroperator}) by making the (rather obvious) following replacements:
\begin{equation}
\hat{\alpha}_1^a \to \alpha_1^a \qq \hat{\alpha}_2^a \to \alpha_2^a \qq \alpha_1^c \to \hat{\alpha}_1^c \qq \alpha_2^c \to \hat{\alpha}_2^c \qq 
\alpha_1^d \to \hat{\alpha}_1^d \qq \alpha_2^d \to \hat{\alpha}_2^d \qq E_p \to (-E_{\hat p}) \qq E_q \to (-E_{\hat q}) \,.
\label{eq:replace-antipode}
\end{equation}
We will consider the consequences of the existence of all these channels with respect to the apparition of new thresholds for the two-body decay process in the next sections.

\section{Threshold analysis}

\subsection{Ultrarelativistic limit of the MKE}
\label{UR}

As we explained in the Introduction, the modifications in the kinematics can generate thresholds (absent in the special-relativistic theory) at energies much lower than the scale $\Lambda$ when the dominant corrections in the ultrarelativistic (UR) limit are different from zero.\footnote{It was mentioned in the Introduction that in theories where the Lorentz symmetry is deformed instead of broken, that is, when the modified kinematics contains a relativity principle, a decay cannot have a threshold (exactly as happens in SR). In fact in these situations the dominant corrections to any kinematic process are zero, so that no new thresholds or relevant corrections to thresholds present in SR are generated by the new kinematics. This will be explicitely proved in the context of a non-universal deformation of SR in Ref.~\cite{Carmona:2014aba}.} This phenomenologically interesting scenario is what we will assume in the present paper.

Let us consider, for definiteness, the MKE corresponding to the $\hat{k}\oplus p \oplus q=0$, Eq.~\eqref{eq:master}. In the UR limit we have
\begin{equation}
\vec{p}\cdot\vec{q}=pq\cos\theta\approx \left(E_p E_q-\frac{m_c^2 E_q}{2E_p}-\frac{m_d^2 E_p}{2 E_q}\right)\cos\theta\,,
\label{eq:URpq}
\end{equation}
where $\theta$ is the angle between the two out-going three momenta. Introducing the adimensional varible $x$, $-1< x < 1$,
\begin{equation}
x\equiv \frac{E_p-E_q}{E_p+E_q} \qq \Leftrightarrow \qq E_p=\frac{1+x}{2}(E_p+E_q)\,, \qq E_q=\frac{1-x}{2}(E_p+E_q)\,,
\label{eq:x-def}
\end{equation}
we can write the UR limit of Eq.~\eqref{eq:masteroperator}, $\hat{O}_3^A$, (which means neglecting all terms containing masses) as
\begin{equation}
\hat{O}_3^A\approx \frac{(E_p+E_q)^3}{\Lambda}\hat{\xi}_3^A(x)\,,
\label{eq:URmasterop}
\end{equation}
where
\begin{equation}
\hat{\xi}_3^A(x)=\eta^c\left(\frac{1+x}{2}\right)^2+\eta^d\left(\frac{1-x}{2}\right)^2+\hat{\eta}^a+2\eta^{ac}\left(\frac{1+x}{2}\right)+2\eta^{ad}\left(\frac{1-x}{2}\right)-2\eta^{cd}\left(\frac{1+x}{2}\right)\left(\frac{1-x}{2}\right)\,,
\label{eq:xiA}
\end{equation}
and we have made the definitions
\begin{equation}
\eta^m\equiv \alpha_1^m+\alpha_2^m \qq \hat{\eta}^m \equiv \hat{\alpha}_1^m+\hat{\alpha}_2^m \qq \eta^{mn}\equiv \beta_1^{mn}+\beta_2^{mn}-\gamma_1^{mn}-\gamma_2^{mn}\,,
\label{eq:etadef}
\end{equation}
where the superscripts $m,n$ stand for $a$, $b$ or $c$ (indicating the particle type).

The equation describing the kinematics of the decay, Eq.~\eqref{eq:master}, in the UR limit, can then be written as
\begin{equation}
\frac{(1+x)(1-x)}{2}(E_p+E_q)^2(1-\cos\theta) + \left(\frac{1-x}{1+x}m_c^2+\frac{1+x}{1-x}m_d^2\right)\cos\theta=m_a^2-m_c^2-m_d^2+\hat{\xi}_3^A(x)\frac{(E_p+E_q)^3}{\Lambda}\,.
\label{eq:UR-MKE}
\end{equation}
One can see from the previous equation that in order for the correction $\hat{O}_3^A$ to be able to generate a threshold in an allowed decay in SR it is necessary that $\hat{\xi}_3^A<0$ (so that it gives a negative contribution to $m_a^2-m_c^2-m_d^2$ which is positive). On the contrary, a kinematically forbidden decay in SR may be allowed above a certain energy threshold if $\hat{\xi}_3^A>0$. It is also obvious that the last term in the right-hand side of Eq.~\eqref{eq:UR-MKE} will not be negligible in the equation in a situation in which it is comparable to the mass terms, something which happens for energies of the order $E\sim (m^2 \Lambda)^{1/3}$ (this gives us the order of magnitude of the threshold energy).\footnote{To get some feeling about this energy scale we can note that $(\Lambda m_a^2)^{1/3}=2.15\left[(\Lambda/M_P)(m_a/\text{GeV})^2\right]^{1/3}\,\text{PeV}$.} 
We will show in the Appendix that in the two cases explained above (allowed and forbidden decay in SR), the threshold of the process is reached exactly for $\cos\theta=1$, so that one can neglect the mass independent term in the left hand side of Eq.~\eqref{eq:UR-MKE}.

\subsection{Threshold determination for a single channel}
\label{thresholddeterm}

According to the comments of the previous subsection, in order to determine the threshold energy we can take $\cos\theta=1$ in the UR limit of the MKE, Eq.~\eqref{eq:UR-MKE}. This gives
\begin{equation}
\frac{2 m_c^2}{1+x}+\frac{2 m_d^2}{1-x}-m_a^2 = \hat{\xi}_3^A (x) \,\frac{(E_p+E_q)^3}{\Lambda} \,,
\label{eq:UR-MKE-notheta}
\end{equation}
which is a relation between $(E_p+E_q)$ and the variable $x$ defined in Eq.~\eqref{eq:x-def}. To determine the value of $x$ producing the maximum (minimum) value of $(E_p+E_q)$ when $\hat{\xi}_3^A<0$ ($\hat{\xi}_3^A>0$), which corresponds to a threshold in an allowed (forbidden) decay in SR, we can derive Eq.~\eqref{eq:UR-MKE-notheta} taking it as an implicit equation which determines $(E_p+E_q)$ as a function of $x$, which gives
\begin{equation}
-\frac{2 m_c^2}{(1+x)^2}+\frac{2 m_d^2}{(1-x)^2}=\frac{d\hat{\xi}_3^A (x)}{d x}\frac{(E_p+E_q)^3}{\Lambda}+3\hat{\xi}_3^A(x)\frac{(E_p+E_q)^2}{\Lambda}\frac{d(E_p+E_q)}{dx} \,.
\label{eq:derivative-x}
\end{equation}
The threshold will correspond to $d(E_p+E_q)/dx=0$, so that from the previous relation we get\footnote{\label{check-xi-0}This is so if $\hat{\xi}_3^A(x)\neq 0$. Therefore, after obtaining $x$ at the threshold, one must check this consistency condition.}
\begin{equation}
-\frac{2 m_c^2}{(1+x)^2}+\frac{2 m_d^2}{(1-x)^2}=\frac{d\hat{\xi}_3^A (x)}{d x}\frac{(E_p+E_q)^3}{\Lambda} \,.
\label{eq:derivative-x-2}
\end{equation}
When one combines Eq.~\eqref{eq:derivative-x-2} with Eq.~\eqref{eq:UR-MKE-notheta}, one gets
\begin{equation}
\left(\frac{2 m_c^2}{1+x}+\frac{2 m_d^2}{1-x}-m_a^2\right)\frac{d\hat{\xi}_3^A}{d x}= \left(-\frac{2 m_c^2}{(1+x)^2}+\frac{2 m_d^2}{(1-x)^2}\right)\hat{\xi}_3^A(x)\,.
\label{eq:threshold-1}
\end{equation}
If we divide now this equation by $m_a^2$, introducing the adimensional quotients
\begin{equation}
\tau_c=\frac{m_c^2}{m_a^2}\qq \qq \tau_d=\frac{m_d^2}{m_a^2}\,,
\label{eq:taudef}
\end{equation}
and we multiply by $(1+x)^2(1-x)^2$, we get a polynomial equation (in general, of order 5 taking into account the general expression for $\hat{\xi}_3^A$ in Eq.~\eqref{eq:xiA}) for $x$:
\begin{equation}
(1-x^2)\left[2(\tau_d+\tau_c)-1+2(\tau_d-\tau_c)x+x^2\right]\frac{d\hat{\xi}_3^A}{d x}=2\left[(\tau_d-\tau_c)+2(\tau_d+\tau_c)x+(\tau_d-\tau_c)x^2\right]\hat{\xi}_3^A\,.
\label{eq:threshold-2}
\end{equation}

The real-valued solutions of Eq.~\eqref{eq:threshold-2} with $-1< x < 1$ give relative maxima or minima of the total energy and
are candidates to the value of $x$ at the threshold. Once determined $x$, the corresponding value of $(E_p+E_q)$, $(E_p+E_q)_*$, can be obtained from Eq.~\eqref{eq:UR-MKE-notheta} or alternatively from Eq.~\eqref{eq:derivative-x-2}, giving
\begin{equation}
(E_p+E_q)_*=(\Lambda m_a^2)^{1/3}\left[\frac{\left(\frac{2\tau_c}{1+x}+\frac{2\tau_d}{1-x}-1\right)}{\hat{\xi}_3^A(x)}\right]^{1/3}=
(\Lambda m_a^2)^{1/3}\left[\frac{2\left(\frac{\tau_d}{(1-x)^2}-\frac{\tau_c}{(1+x)^2}\right)}{d\hat{\xi}_3^A/dx}\right]^{1/3}.
\label{eq:threshold-en}
\end{equation}

If Eq.~\eqref{eq:threshold-2} has several real-valued solutions for $x$ in the interval $-1< x < 1$, then, in order to determine the actual threshold, one needs to calculate their corresponding $(E_p+E_q)_*$ according to Eq.~\eqref{eq:threshold-en}. In the case of an allowed decay in SR (and with $\hat{\xi}_3^A<0$ in order to generate a threshold) the largest of the $(E_p+E_q)_*$ values is the threshold of the reaction (since the corresponding value of $x$ gives a configuration along which the reaction can still proceed). For higher energies the MKE of the reaction has no solution and therefore the decay becomes kinematically forbidden. In the case of a decay which is forbidden in SR (and with $\hat{\xi}_3^A>0$ in order to generate a threshold) the smallest of the $(E_p+E_q)_*$ values giving solutions to the MKE corresponds to an energy below which the equation has no solution and therefore the decay is kinematically forbidden. In the case $\tau_d=0$ (or $\tau_c=0$) one has also to compare the obtained value of $(E_p+E_q)_*$ with the value that $(E_p+E_q)$ takes when $x\to 1$ (or $x\to -1$) in order to determine the supremum or infimum of the total energy in the interval $-1< x < 1$.

Note that the threshold energy $(E_p+E_q)_*$ defined by Eq.~\eqref{eq:threshold-en} is not in fact the energy of the decaying particle at the threshold. However, all the energies involved are much smaller than $\Lambda$, a necessary condition since our model only considers first order corrections to the standard kinematics in an expansion in powers of $1/\Lambda$. Therefore, in the regime of applicability of the present model, the difference between the value of the energy of the decaying particle at the threshold and $(E_p+E_q)_*$ will be of order $(E_p+E_q)_*^2/\Lambda\ll (E_p+E_q)_*$. This will be also the case for energy or momentum measurements much lower than the scale $\Lambda$. The differences in their determination by different measurements methods (owing to, for example, the uncertainty in the channel through which a specific process took place) will be much smaller than the experimental errors in the corresponding measurements. Of course, our model becomes useless when the energy approaches the scale $\Lambda$ of new physics, but this is exactly the situation of an effective field theory, that in our case is realized at a kinematic level.

\subsection{Threshold for the multichannel kinematics}

We argued in Section~\ref{sec:kinematics} that we have twelve different channels, corresponding to the possible generalizations of the usual energy-momentum conservation laws, along which the decay can proceed. For each of them we can write the UV limit of the correction proportional to $1/\Lambda$ in the MKE as
\begin{equation}
O_3\approx \frac{(E_p+E_q)^3}{\Lambda}\xi_3(x)\,,
\label{eq:general-xi}
\end{equation}
where $\xi_3(x)$ can be obtained from $\hat{\xi}_3^A$ in Eq.~\eqref{eq:xiA} by appropriate substitutions for the different channels, such as those indicated in Eqs.~\eqref{eq:replacechannel-1}, \eqref{eq:replacechannel-2}, or \eqref{eq:replace-antipode}. We will write the explicit expressions in the next Section for the specific case of Cherenkov radiation.

The decay process $A\to C+D$ can proceed through any kinematic configuration $(\vec{p},\vec{q})$ which satisfies any of the MKE of the different channels. Therefore, in the case of an allowed decay in SR, the threshold energy generated by the new kinematics is the largest of the threshold energies calculated according to the previous subsection for all the different channels. Similarly, in the case of a forbidden decay in SR, the threshold energy giving the energy above which the decay is allowed with the new kinematics will correspond to the minimum of the threshold energies corresponding to each of the channels.

The application of this method to determine the threshold induced by the kinematics beyond SR for any particle process is crucial to put correct phenomenological bounds on the $(\alpha,\beta,\gamma)$ coefficients of the MDR and the MCL which define the new kinematics, and, up to our knowledge, has never done before in the literature. In the following section we will do a explicit calculation in the case of a well-known process in the context of Lorentz violating kinematics: vacuum Cherenkov radiation. 

\section{Bounds on vacuum Cherenkov radiation}
\label{sec:Cherenkov}

We will apply the formalism developed in the previous Sections to the case of the decay $e^-\to e^-+\gamma$ in vacuum, a process forbidden by SR. The study of the kinematic extensions of SR which are excluded by present phenomenological bounds have been carried out in previous works by applying the condition that there can be no threshold in the previous process for electron energies below $50$\,TeV. This seems to be a necessary condition to reproduce the observed radiation coming from the Crab Nebula, since this is the expected energy of the electrons responsible for the creation, via inverse Compton scattering, of the more energetic photons we observe from that source~\cite{Mattingly:2005re}.

However, these studies were made under the assumption that there are no modifications on the usual conservation laws. This may be too naive, as we argued in the Introduction, and our main objective in this Section will be to show that such bounds may be incorrect in the more general case of modified composition laws. Of course, this has relevant consequences when one uses these bounds to estimate other phenomenological aspects of Planckian physics, such as effects in other particle processes, in experiments of time-of-flight delays, etc.

Our approach will be purely kinematical, since, as we remarked in the Introduction, in contrast to what happens in the effective field theory framework, we do not have a dynamic theory at our disposal. A possible criticism, then, is that the dynamical matrix element could affect the limits extracted from just kinematic considerations. However,
under the assumption that the dynamics does not change drastically from that of the Lorentz invariant theory, one can estimate by simple dimensional analysis that the decay time will be extremely short above threshold, $\tau\sim\Lambda/(50\,\text{TeV})^2\sim 10^{-15}\,$s, which in our case will impede the formation of a population of electrons at the energies necessary to produce the observed $50\,\text{TeV}$ photons. This is a general conclusion which has been applied also in the case of previous studies in the framework of effective field theory: for rapid reactions (it would not be the case of decays involving the weak interaction, or for thresholds in certain scattering processes) one can forget about the dynamics and derive constraints using only kinematic models~\cite{Mattingly:2005re}.

\subsection{Multichannel analysis of the process}

In the case of $e^-(k)\to e^-(p)+\gamma(q)$, Eq.~\eqref{eq:xiA} gets simplified, since in this case the particle types $(a)$ and $(c)$ coincide. We will make a change of notation accordingly:
\begin{equation}
\eta^c=\eta^e \qq \hat{\eta}^a=\hat{\eta}^e \qq \eta^d=\eta^\gamma \qq \eta^{ad}=\eta^{cd}=\eta^{e\gamma} 
\label{eq:notation-1}
\end{equation}
and
\begin{equation}
\eta^{ac}=\eta^{ee}=\beta_1^e+\beta_2^e-\gamma_1^e-\gamma_2^e=-\frac{\hat{\eta}^e+\eta^e}{2}\,,
\label{eq:notation-2}
\end{equation}
where in the last equation we have used the definition of $\eta^{mn}$, Eq.~\eqref{eq:etadef}, and Eq.~\eqref{eq:coefMDRantipode}. The expression of $\hat{\xi}_3^A$ in Eq.~\eqref{eq:xiA} is then
\begin{align}
\hat{\xi}_3^A&=\eta^e\left(\frac{1+x}{2}\right)^2+\eta^\gamma\left(\frac{1-x}{2}\right)^2+\hat{\eta}^e+\eta^{ee}(1+x)+\eta^{e\gamma}(1-x)-\eta^{e\gamma}(1-x)\left(\frac{1+x}{2}\right) \nonumber \\ &=
(1-x)\left[-\eta^{ee}-\eta^e+\frac{1}{2}\left(\eta^{e\gamma}+\frac{\eta^e+\eta^\gamma}{2}\right)(1-x)\right]\,,
\label{eq:new-xiA}
\end{align}
where we have used Eqs.~\eqref{eq:notation-1}-\eqref{eq:notation-2} and regrouped terms taking into account the identity $(1+x)/2=1-(1-x)/2$.

As we explained in the previous sections, there are twelve different channels corresponding to the different forms that the conservation law can adopt; these are classified in Table~\ref{table:channels}. From the substitutions in the coefficients needed to write a MKE for a channel from that of a different channel explained in Section~\ref{sec:comments}, it is obvious that all the $\xi_3(x)$ in Eq.~\eqref{eq:general-xi} will have a similar form to that of $\hat{\xi}_3^A$ in Eq.~\eqref{eq:new-xiA}, that is:
\begin{equation}
\xi_3(x)=(1-x)\left[a_0+\frac{a_1}{2}(1-x)\right].
\label{eq:general-xi-expression}
\end{equation}

\begin{table}%
\begin{tabular}{|c|c|c|c|c|c|}\hline
$C_1$: & $\hat{k}\oplus p \oplus q = 0$ & $C_2$: & $\hat{k}\oplus q \oplus p = 0$ & $C_3$: & $p \oplus \hat{k}\oplus q = 0$  \\ \hline
$C_4$: & $q\oplus\hat{k}\oplus p = 0$ & $C_5$: & $p\oplus q\oplus\hat{k} = 0$ & $C_6$: & $q \oplus p \oplus \hat{k} = 0$  \\ \hline
$C_7$: & $ k \oplus\hat{p}\oplus\hat{q}= 0$ & $C_8$: & $ k \oplus\hat{q}\oplus\hat{p}= 0$ & $C_9$: & $\hat{p} \oplus  k \oplus\hat{q}= 0$  \\ \hline
$C_{10}$: & $\hat{q}\oplus k \oplus\hat{p}= 0$ & $C_{11}$: & $\hat{p}\oplus \hat{q}\oplus k  = 0$ & $C_{12}$: & $\hat{q} \oplus\hat{p}\oplus  k  = 0$  \\ \hline
\end{tabular}
\caption{Notation to design the different channels in the decay $A(k)\to C(p)+D(q)$.}
\label{table:channels}
\end{table}

The coefficients $a_0$, $a_1$ are linear combinations of the adimensional coefficients in the MDR $(\eta^e,\eta^\gamma)$ and in the MCL $(\eta^{ee},\eta^{e\gamma},\eta^{\gamma e},\eta^{\gamma\gamma})$, which are different for each of the channels. They are given in Table~\ref{table:coef-channels}, where conveniently we have introduced the combinations of parameters
\begin{equation}
\lambda^{ee}=\eta^{ee}+\eta^e \qq \lambda^{\gamma\gamma}=\eta^{\gamma\gamma}+\eta^\gamma \qq
\lambda^{e\gamma}=\eta^{e\gamma}+\frac{\eta^e+\eta^\gamma}{2} \qq
\lambda^{\gamma e}=\eta^{\gamma e}+\frac{\eta^\gamma+\eta^e}{2} \,.
\label{eq:def-lambda}
\end{equation}

\begin{table}%
\begin{tabular}{|c|c|c|}\hline
$C_1, C_3$: & $a_0=-\lambda^{ee}$ & $a_1=\lambda^{e\gamma}$  \\ \hline
$C_2$: & $a_0=\lambda^{e\gamma}-\lambda^{\gamma e}-\lambda^{ee}$ & $a_1=\lambda^{\gamma e}$  \\ \hline
$C_4, C_6$: & $a_0=-\lambda^{ee}$ & $a_1=\lambda^{\gamma e}$  \\ \hline
$C_5$: & $a_0=\lambda^{\gamma e}-\lambda^{e\gamma}-\lambda^{ee}$ & $a_1=\lambda^{e\gamma}$  \\ \hline
$C_7, C_9$: & $a_0=-\lambda^{ee}$ & $a_1=(\lambda^{ee}+\lambda^{\gamma\gamma})-\lambda^{e\gamma}$  \\ \hline
$C_8$: & $a_0=\lambda^{\gamma e}-\lambda^{e\gamma}-\lambda^{ee}$ & $a_1=(\lambda^{ee}+\lambda^{\gamma\gamma})-\lambda^{\gamma e}$  \\ \hline
$C_{10}, C_{12}$: & $a_0=-\lambda^{ee}$ & $a_1=(\lambda^{ee}+\lambda^{\gamma\gamma})-\lambda^{\gamma e}$  \\ \hline
$C_{11}$: & $a_0=\lambda^{e\gamma}-\lambda^{\gamma e}-\lambda^{ee}$ & $a_1=(\lambda^{ee}+\lambda^{\gamma\gamma})-\lambda^{e\gamma}$  \\ \hline
\end{tabular}
\caption{Values of the coefficients of Eq.~\eqref{eq:general-xi-expression} for the channels $C_1$-$C_{12}$ defined in Table~\ref{table:channels}. The coefficients $\lambda^{mn}$ are defined in Eq.~\eqref{eq:def-lambda}.}
\label{table:coef-channels}
\end{table}

In summary, from the 24 parameters that generically characterize the MDR and MCL for the electron-photon system $(\{\alpha_1^e,\alpha_2^e\},\{\alpha_1^\gamma,\alpha_2^\gamma\},\{\beta_1^e,\beta_2^e,\gamma_1^e,\gamma_2^e,\gamma_3^e\},\{\beta_1^\gamma,\beta_2^\gamma,\gamma_1^\gamma,\gamma_2^\gamma,\gamma_3^\gamma\},\{\beta_1^{e\gamma},\beta_2^{e\gamma},\gamma_1^{e\gamma},\gamma_2^{e\gamma},\gamma_3^{e\gamma}\},\{\beta_1^{\gamma e},\beta_2^{\gamma e},\gamma_1^{\gamma e},\gamma_2^{\gamma e},\gamma_3^{\gamma e}\})$, only the four combinations $(\lambda^{ee},\lambda^{\gamma\gamma},\lambda^{e\gamma},\lambda^{\gamma e})$ defined in Eq.~\eqref{eq:def-lambda}\footnote{Recall that the six combinations of parameters $(\eta^e,\eta^\gamma,\eta^{ee},\eta^{\gamma\gamma},\eta^{e\gamma},\eta^{\gamma e})$ were defined in Eq.~\eqref{eq:etadef}.} are relevant for a multichannel discussion of the threshold of the process. 

\subsection{Excluded regions for a generic channel}
\label{sec:excluded-generic}
  
We will now apply the threshold analysis that was described in Section~\ref{thresholddeterm} for a generic channel. We start from Eq.~\eqref{eq:general-xi-expression}, where the $a_0$ and $a_1$ coefficients appearing there are different functions of the four combinations of parameters $(\lambda^{ee},\lambda^{\gamma\gamma},\lambda^{e\gamma},\lambda^{\gamma e})$ for the different channels, according to Table~\ref{table:coef-channels}.

The value of $x$ at the threshold is either a solution of Eq.~\eqref{eq:threshold-2} for $-1<x<1$, which in this case $(\tau_c=1,\tau_d=0)$ reduces to
\begin{equation}
(1-x^2)\frac{d\xi_3}{d x}=-2\xi_3(x)\,,
\label{eq:x-equation}
\end{equation}
or $x=1$.
From Eq.~\eqref{eq:general-xi-expression}, $d\xi_3/d x=-d\xi_3/d (1-x)=-a_0-a_1(1-x)$, so that Eq.~\eqref{eq:x-equation} gives
\begin{equation}
(1-x^2)[a_0+a_1(1-x)]=2(1-x)\left[a_0+\frac{a_1}{2}(1-x)\right]\,,
\label{eq:x-equation-2}
\end{equation}
whose solution is just $x_*=a_0/a_1$. The corresponding value of the total energy is obtained with the help of Eq.~\eqref{eq:threshold-en}:
\begin{equation}
(E_p+E_q)_*=(\Lambda m_e^2)^{1/3}\left[\frac{2 a_1}{(a_0+a_1)^2}\right]^{1/3}\,.
\label{eq:Ch-threshold}
\end{equation}
In order for Eq.~\eqref{eq:Ch-threshold} to be a valid candidate to the threshold we must impose that $x_*$ be a number between $-1$ and $1$, $-a_1<a_0<a_1$, and positivity of the energy $E_*$, that is: $a_1>0$. 

Finally, as explained at the end of Sec.~\ref{thresholddeterm}, since $\tau_d=0$, the limit of $(E_p+E_q)$ when $x\to 1$ is finite, so that the correct threshold of the process will be given by the minimum between Eq.~\eqref{eq:Ch-threshold} and the value of ($E_p+E_q$) at $x=1$. If we go back to Eq.~\eqref{eq:UR-MKE-notheta} and particularize it to $m_c=m_a=m_e$, $m_d=0$, it gives
\begin{equation}
\frac{m_e^2}{1+x}=\left[a_0+\frac{a_1}{2}(1-x)\right]\frac{(E_p+E_q)^3}{\Lambda}\,.
\label{eq:new-UR-MKE-notheta}
\end{equation}
Then we have
\begin{equation}
(E_p+E_q)(x=1)=(\Lambda m_e^2)^{1/3}\left[\frac{1}{2 a_0}\right]^{1/3}\,.
\label{eq:Ch-threshold2}
\end{equation}
It will be a valid candidate for the threshold only when $a_0>0$, since the energy must be positive. However, it is simple to see that for any $a_0>0, a_1>0$, the value of $(E_p+E_q)_*$ from Eq.~\eqref{eq:Ch-threshold} is always lower than $(E_p+E_q)(x=1)$ from Eq.~\eqref{eq:Ch-threshold2}. Just note that
\begin{equation}
(a_1-a_0)^2 >0 \quad\Rightarrow\quad a_1^2+a_0^2-2a_1 a_0>0 \quad\Rightarrow\quad (a_1+a_0)^2 > 4 a_1a_0 \quad\Rightarrow\quad \frac{2 a_1}{(a_1+a_0)^2}<\frac{1}{2 a_0}\,.
\label{eq:inequalities}
\end{equation}
Therefore, in the region where \eqref{eq:Ch-threshold} is a valid solution, this will be the correct value of the threshold. Outside this region, and as long as $a_0>0$, the correct threshold will be given by Eq.~\eqref{eq:Ch-threshold2}. If $a_0<0$ and $a_0<-a_1$, then there is no threshold and Cherenkov radiation will not be produced.

\begin{figure}
\centerline{\includegraphics[scale=0.85]{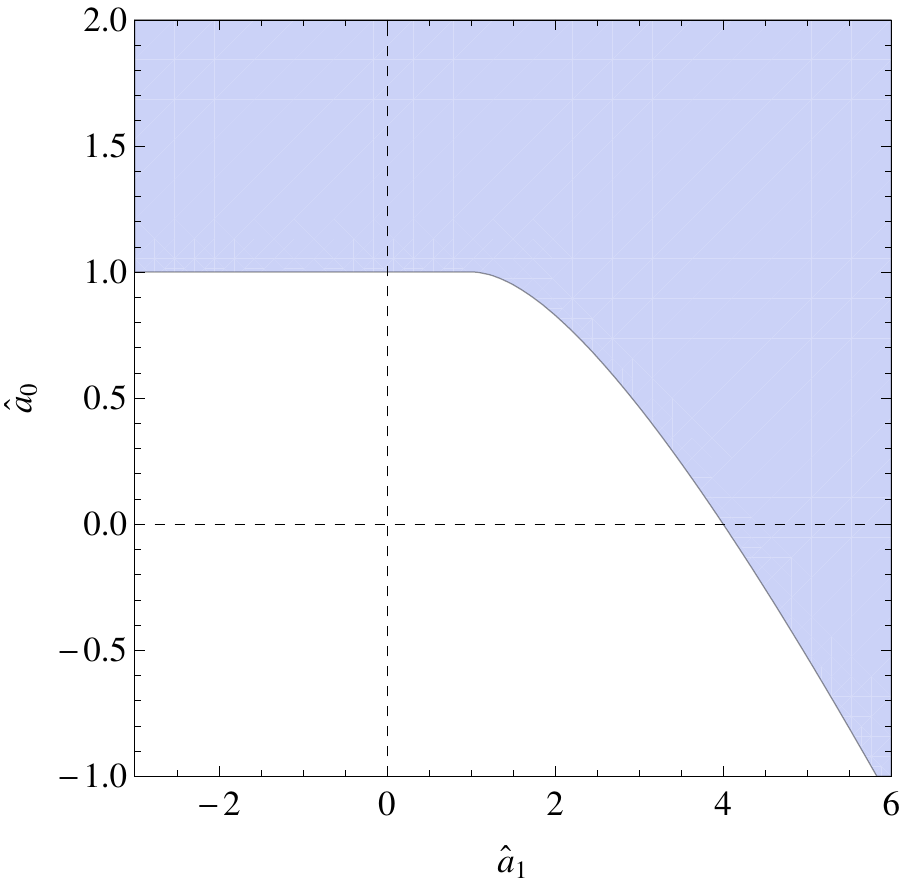}}
\caption{Excluded values of $(\hat{a}_1,\hat{a}_0)$ represented as a blue solid region by the phenomenological limits on the production of Cherenkov radiation.}
\label{fig:a1a0}
\end{figure}

%In this case, vacuum Cherenkov radiation will be produced above the threshold Eq.~\eqref{eq:Ch-threshold}. 
Since, as we said at the beginning of this Section, one should exclude phenomenologically electron decays below $E_*<50\,$TeV, we have the following excluded region of the $(a_0,a_1)$ parameter space:
\begin{equation}
R_1: \qq a_1>0 \qq -a_1<a_0<a_1 \qq \frac{(a_0+a_1)^2}{2 a_1}> 2.5\times 10^{-2}\,\left(\frac{\Lambda}{M_P}\right)\,,
\label{eq:excluded}
\end{equation}
together with 
\begin{align}
R_2: \qq & a_1>0 \qq  a_0>a_1 \qq  2 a_0 > 2.5\times 10^{-2}\,\left(\frac{\Lambda}{M_P}\right)\,, \\
R_3: \qq & a_1<0 \qq a_0>0 \qq  2 a_0 > 2.5\times 10^{-2}\,\left(\frac{\Lambda}{M_P}\right)\,.
\label{eq:excluded-2}
\end{align}
The above excluded regions can be expressed in a simpler way in terms of the variables
\begin{equation}
\hat{a}_0\equiv \frac{a_0}{b} \qq \hat{a}_1\equiv \frac{a_1}{b}\,,
\label{eq:normaliz}
\end{equation} 
where we have introduced the constant
\begin{equation}
b\equiv 1.25\times 10^{-2}\,\left(\frac{\Lambda}{M_P}\right)\,.
\label{eq:constant-b}
\end{equation}
The region $R_2$ can then be written as $\hat{a}_1>0$, $\hat{a}_0>\hat{a}_1$, $\hat{a}_0>1$, while the region $R_3$ is simply $\hat{a}_1<0$, $\hat{a}_0>1$. On the other hand, the region $R_1$ can be written as $\hat{a}_1>0$, $-\hat{a}_1<\hat{a}_0<\hat{a}_1$, $(\hat{a}_0+\hat{a}_1)^2>4\hat{a}_1$, which is equivalent to $\hat{a}_1>0$, $\hat{a}_0<\hat{a}_1$, $\hat{a}_0>-\hat{a}_1+2\sqrt{\hat{a}_1}$, or $\hat{a}_1>0$, $-\hat{a}_1+2\sqrt{\hat{a}_1}<\hat{a}_0<\hat{a}_1$. Note that region $R_1$ only exists if $-\hat{a}_1+2\sqrt{\hat{a}_1}<\hat{a}_1$, which implies that $\hat{a}_1>1$. Therefore, for $\hat{a}_1>1$, the excluded regions $R_1$ and $R_2$ form a single region determined by $\hat{a}_0>-\hat{a}_1+2\sqrt{\hat{a}_1}$. In summary, the production of Cherenkov radiation excludes the set of points in the parameter space $(\hat{a}_1,\hat{a}_0)$ which lie above the curve
\begin{equation}
\left\{\begin{array}{ll}
\hat{a}_0=1 \quad & \text{for}\quad \hat{a}_1<1 \\
\hat{a}_0=-\hat{a}_1+2\sqrt{\hat{a}_1} \quad & \text{for} \quad \hat{a}_1>1
\end{array}\right.
\label{eq:exclusion-curve}
\end{equation} 
This set of points is represented as a blue solid region in Fig.~\ref{fig:a1a0}.

This excluded region should now be translated in excluded regions in the parameter space $(\lambda^{ee},\lambda^{\gamma\gamma},\lambda^{e\gamma},\lambda^{\gamma e})$ for each of the possible channels. The intersection of these regions is the true excluded region in the parameter space, since as long as one point in the parameter space is phenomenologically viable for one of the channels, one cannot discard that the decays proceeds through that channel. The multichannel excluded region would then translate into bounds for the six parameters $(\eta^{e},\eta^{\gamma},\eta^{ee},\eta^{\gamma\gamma},\eta^{e\gamma},\eta^{\gamma e})$ defining the modified kinematics. In terms of these parameters, we see that there is a two-dimensional degeneracy in the bounds we can get from Cherenkov radiation: there are different values of the six parameters that produce identical phenomenological effect with respect to Cherenkov radiation.

\subsection{Excluded regions for particular cases of the modified kinematics}

Since the general discussion is rather involved, we will consider here some particular cases of the modified kinematics in which one can extract useful physical conclusions.

\begin{figure}
\centerline{\includegraphics[scale=0.85]{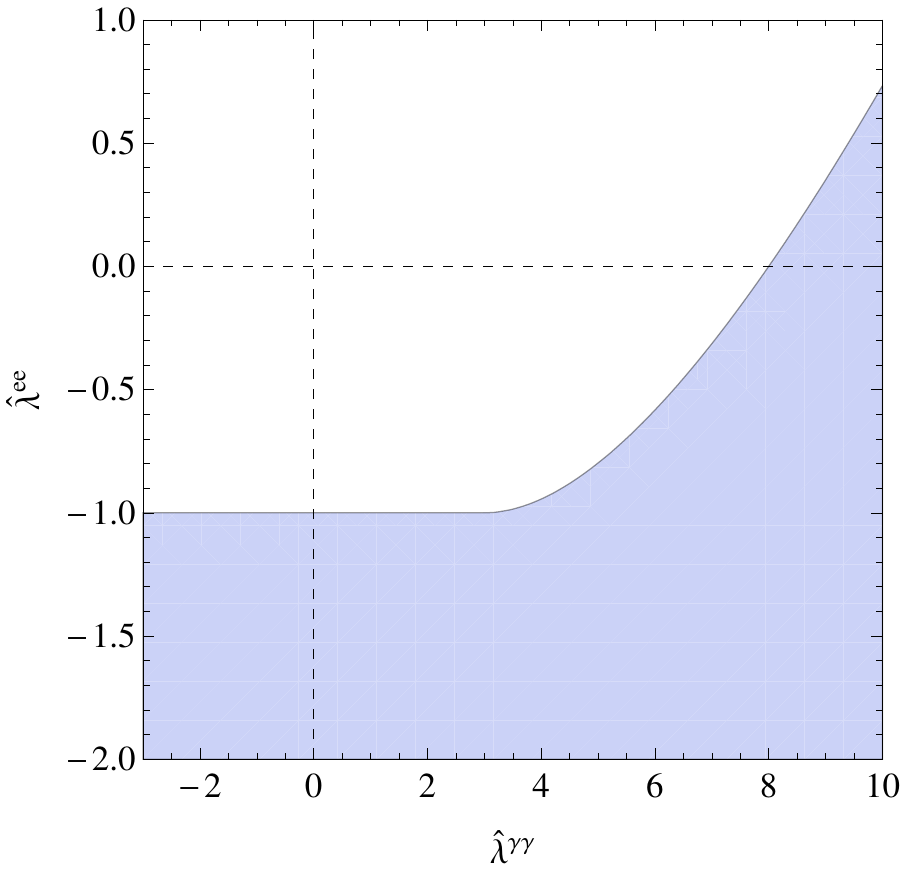} \qq \qq \qq \includegraphics[scale=0.85]{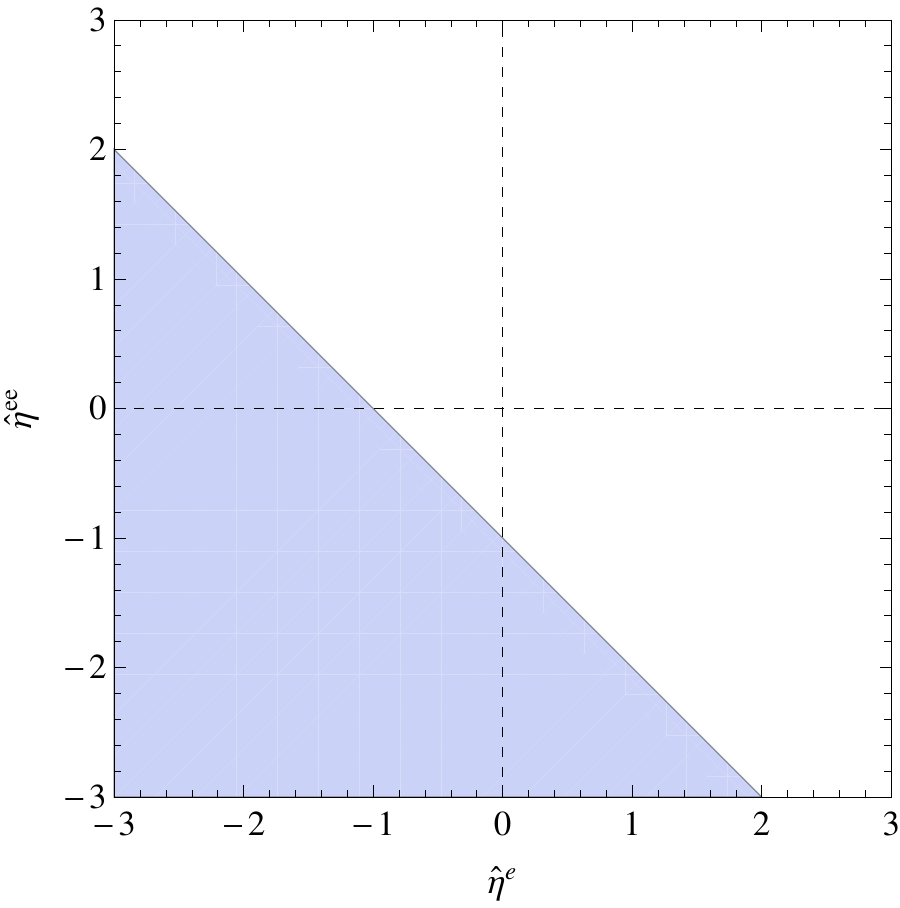}}
\caption{Excluded regions in: (left panel) the $(\hat{\lambda}^{\gamma\gamma},\hat{\lambda}^{ee})$ plane for the case given by Eq.~\eqref{eq:one-channel}; (right panel) the $(\hat{\eta}^e,\hat{\eta}^{ee})$ plane for the case given by Eq.~\eqref{eq:universal-channel}.}
\label{fig:one-channel}
\end{figure}

As we have seen, in the general case there are twelve channels that determine an excluded region in the four-dimensional $(\lambda^{ee},\lambda^{\gamma\gamma},\lambda^{e\gamma},\lambda^{\gamma e})$ parameter space, and which translates into bounds into the six coefficients $(\eta^{e},\eta^{\gamma},\eta^{ee},\eta^{\gamma\gamma},\eta^{e\gamma},\eta^{\gamma e})$. The analysis simplifies greatly in the following particular cases:

\begin{enumerate}
\item $\eta^{e \gamma}=\eta^{\gamma e}$. In this case there are 5 independent parameters defining the modified kinematics: $(\eta^{e},\eta^{\gamma},\eta^{ee},\eta^{\gamma\gamma},\eta^{e\gamma})$, which translate into three independent combinations of parameters $(\lambda^{ee},\lambda^{\gamma\gamma},\lambda^{e\gamma})$, since $\lambda^{\gamma e}=\lambda^{e\gamma}$. From Table~\ref{table:coef-channels} we can see that in this case there are only two independent channels:
\begin{align}
C_1=...=C_6 \qq a_0&=-(\eta^{ee}+\eta^e) \qq a_1=\eta^{e\gamma} + \frac{\eta^e+\eta^\gamma}{2} \nonumber \\
C_7=...=C_{12} \qq a_0&=-(\eta^{ee}+\eta^e) \qq a_1=\eta^{ee}+\eta^{\gamma\gamma}-\eta^{e\gamma} + \frac{\eta^e+\eta^\gamma}{2}\,.
\label{eq:two-channels}
\end{align}
Fig.~\ref{fig:a1a0} would then define a excluded region in the tridimensional parameter space $(\lambda^{ee},\lambda^{\gamma\gamma},\lambda^{e\gamma})$ or the penta-dimensional parameter space $(\eta^{e},\eta^{\gamma},\eta^{ee},\eta^{\gamma\gamma},\eta^{e\gamma})$.
\item $\eta^{e\gamma}=\eta^{\gamma e}=(\eta^{ee}+\eta^{\gamma\gamma})/2$. This is a subcase of the previous case, in which there are only 4 independent parameters defining the modified kinematics: $(\eta^{e},\eta^{\gamma},\eta^{ee},\eta^{\gamma\gamma})$. They translate into two independent combinations of parameters $(\lambda^{ee},\lambda^{\gamma\gamma})$, since now $\lambda^{e\gamma}=\lambda^{\gamma e}=(\lambda^{ee}+\lambda^{\gamma\gamma})/2$.
In this case there is only one channel, for which
\begin{equation}
a_0=-(\eta^{ee}+\eta^e)\qq a_1=\frac{(\eta^{ee}+\eta^e)}{2}+\frac{(\eta^{\gamma\gamma}+\eta^\gamma)}{2}\,.
\label{eq:one-channel}
\end{equation}
Figure~\ref{fig:one-channel} (left panel) shows the excluded region in the plane $(\hat{\lambda}^{\gamma\gamma},\hat{\lambda}^{ee})\equiv (\lambda^{\gamma\gamma}/b,\lambda^{ee}/b)$.
\item $\eta^\gamma=\eta^e$, $\eta^{\gamma\gamma}=\eta^{e\gamma}=\eta^{\gamma e}=\eta^{ee}$. This is once more a subcase of the previous situation, and corresponds to the particular interesting case of a universal modified kinematics. We have two independent parameters $(\eta^e,\eta^{ee})$ which translate into only one independent combination $\lambda^{ee}$ ($\lambda^{\gamma\gamma}=\lambda^{e\gamma}=\lambda^{\gamma e}=\lambda^{ee}$). There is again one channel for which  
\begin{equation}
a_0=-\lambda^{ee} \qq a_1=\lambda^{ee} \qq\Rightarrow\qq a_1=-a_0\,.
\label{eq:universal-channel}
\end{equation}
We see that in this case, however, $x_*=a_0/a_1=-1$, which is not a valid solution. Therefore, the region $R_1$ does not exist in this case. The excluded region will then be defined simply by $\hat{a}_0>1$, or $\hat{\lambda}^{ee}<-1$. If we define once again $\hat{\eta}^{e}\equiv \eta^e/b$, $\hat{\eta}^{ee}\equiv \eta^{ee}/b$,\footnote{This reescaling is a manifestation of the fact that the modification of the kinematics only depend on the quotients $\eta/\Lambda$.} then the excluded region will be given by $\hat{\eta}^{ee}+\hat{\eta}^e<-1$. This region is showed in Fig.~\ref{fig:one-channel} (right panel).
\end{enumerate}

There exist two other cases particularly interesting: the one in which the dispersion relations of electron and photon are those of SR, but we have modified conservation laws, and the one in which all the modification in the kinematics is made at the level of the MDR, so that the energy-momentum conservation law is that of SR.

In the first case, all the modification to the kinematics of SR is given by new momenta composition laws, so that $\eta^e=\eta^\gamma=0$ and Eq.~\eqref{eq:def-lambda} gives
\begin{equation}
\lambda^{ee}=\eta^{ee} \qq \lambda^{e\gamma}=\eta^{e \gamma} \qq \lambda^{\gamma e}=\eta^{\gamma e} \qq \lambda^{\gamma\gamma}=\eta^{\gamma\gamma}\,.
\label{eq:MCL-threshold}
\end{equation}
As in the general case, in this specific situation there are four independent $\lambda$ combinations and twelve different channels. Therefore this case does not contain any simplification and is in fact equivalent to the general case. This is a manifestation of the degeneracy that we mentioned at the end of Section~\ref{sec:excluded-generic}: the effect of the modified dispersion relations can be reabsorbed by a change in the composition laws, as far as the process of electron decay is concerned. 

The second of the cases referred above ($\eta^{ee}=\eta^{e\gamma}=\eta^{\gamma e}=\eta^{\gamma\gamma}=0$) is the one which has been considered in previous works based on the hypothesis that EFT is a valid framework to explore kinematics beyond SR. In this case there is evidently one channel (since the conservation law is that of SR), which is confirmed by computing
\begin{equation}
\lambda^{ee}=\eta^e \qq \lambda^{e\gamma}=\lambda^{\gamma e}=\frac{\eta^e+\eta^\gamma}{2} \qq \lambda^{\gamma\gamma}=\eta^\gamma\,,
\label{eq:MDR-threshold}
\end{equation}
and checking that the expressions of $a_0$, $a_1$ for the different channels from Table~\ref{table:coef-channels} coincide:
\begin{equation}
a_0=-\eta^e \qq a_1=\frac{\eta^e+\eta^\gamma}{2}\,.
\label{eq:}
\end{equation}

\begin{figure}
\centerline{\includegraphics[scale=0.85]{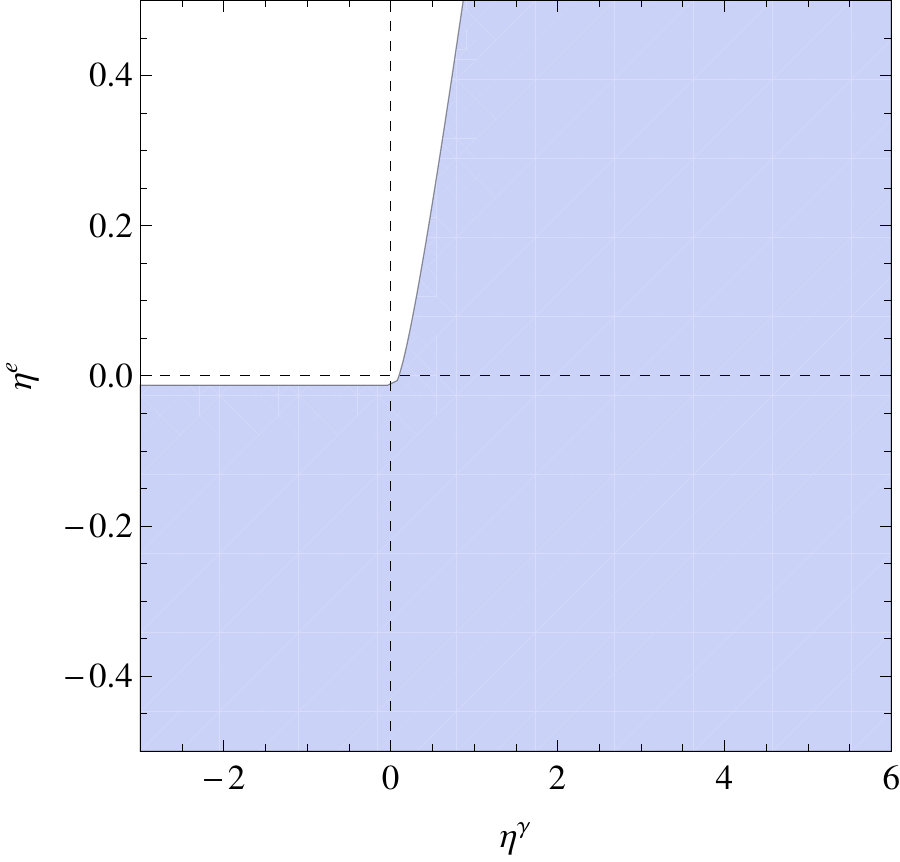}}
\caption{Excluded values of $(\eta^\gamma,\eta^e)$ represented as a blue solid region by the phenomenological limits on the production of Cherenkov radiation in the case where all modification of the kinematics is made at the level of the dispersion relation.}
\label{fig:mdr}
\end{figure}

The excluded region is shown in Figure~\ref{fig:mdr}, where, in order to compare with previous results in the literature, we have used the $(\eta^\gamma,\eta^e)$ instead of the $(\hat{\eta}^\gamma,\hat{\eta}^e)$ plane. We see from that figure that, for example, the values $(\eta^e=0,\eta^\gamma=1)$\footnote{The natural values of the adimensional coefficients in the power expansion of $1/\Lambda$ are of order 1, which is equivalent to say that the modified kinematics is being caused by physics at the $\Lambda$ energy scale.} are excluded when $\Lambda\sim M_P$: present phenomenology is therefore able to exclude certain Planckian effects. This is in fact what had been obtained in previous studies~\cite{Mattingly:2005re}.

We can now ask the following interesting question: are bounds obtained in the EFT framework still valid in a more general situation in which one has modified conservation laws?
Alternatively, are values of coefficients excluded by studies which only considered modifications in dispersion relations also excluded in a more general situation? As we will now show, the answer to this question is no.

Let us take the previous example of the values $(\eta^e=0,\eta^\gamma=1)$, already excluded phenomenologically in the EFT framework. Now let us consider the following point in the six-dimensional parameter space: $(\eta^e=0,\,\eta^\gamma=1,\,\eta^{ee}=0,\,\eta^{\gamma\gamma}=-1,\,\eta^{e\gamma}=-1/2,\,\eta^{\gamma e}=-1/2)$. It is simple to check that in this case $\lambda^{ee}=\lambda^{\gamma\gamma}=\lambda^{e\gamma}=\lambda^{\gamma e}=0$, so that this modified kinematics has no effect at all in vacuum Cherenkov radiation.\footnote{In fact one can see that this choice of parameters is an example of a modified kinematics compatible with a relativity principle~\cite{Carmona:2014aba}.}
 Another example is the point $(\eta^e=0,\,\eta^\gamma=1,\,\eta^{ee}=1,\,\eta^{\gamma\gamma}=-1,\,\eta^{e\gamma}=0,\,\eta^{\gamma e}=0)$. In this case $\lambda^{ee}=1$, $\lambda^{\gamma\gamma}=0$, $\lambda^{e\gamma}=\lambda^{\gamma e}=1/2$. This point in the parameter space is a less trivial example than the previous one, but again the corresponding kinematics does not produce Cherenkov radiation, since $a_0=-1$, $a_1=1/2$, and then the condition $-a_1<a_0<a_1$ is not satisfied. A more interesting example is the point $(\eta^e=0,\,\eta^\gamma=1,\,\eta^{ee}=\sigma/2,\,\eta^{\gamma\gamma}=\sigma/2,\,\eta^{e\gamma}=\sigma/2,\,\eta^{\gamma e}=\sigma/2)$. This is an example of case (ii) of the list above, so that there is only one channel for which, according to Eq.~\eqref{eq:one-channel},
\begin{equation}
a_0=-\frac{\sigma}{2} \qq a_1=\frac{\sigma+1}{2}\,.
\label{eq:example}
\end{equation}
If we take $\sigma>0$, then the conditions $a_1>0$ and $-a_1<a_0<a_1$ are automatically satisfied and then this kinematics will produce Cherenkov radiation above an energy threshold. According to Eq.~\eqref{eq:excluded}, this is still phenomenologically allowed if
\begin{equation}
\frac{(a_0+a_1)^2}{2 a_1}<2.5\times 10^{-2}\,\left(\frac{\Lambda}{M_P}\right) \qq \Leftrightarrow \qq \sigma+1 > 10\,\left(\frac{\Lambda}{M_P}\right)\,.
\label{eq:example2}
\end{equation} 
Taking $\sigma>9$, this is an example of a modified kinematics with Planckian sensitivity which is not excluded by present phenomenological data (but whose Cherenkov radiation prediction could be tested in the near future) and which contains modified dispersion relations which would have been too naively excluded in the EFT framework. This confirms our previous arguments that the EFT framework may be too restrictive concerning Cherenkov decay phenomenology. 

\section{Conclusions}

In this work we have shown how to extend the analysis of thresholds induced by a modification of the dispersion relation of particles when one goes beyond the EFT framework including a modification of the energy-momentum conservation laws. The analysis of thresholds and bounds in this case is much more involved due to the presence of multi-channels. 

A detailed analysis of the bounds on deviations from SR kinematics that can be obtained by considering vacuum Cherenkov radiation has been presented including the results for a few particularly simple cases (universal kinematics, symmetry in the composition of different type of particles).     

There is a degeneracy in the bounds: different values of parameters produce the same effects in vacuum Cherenkov radiation. In order to remove this degeneracy it would be necessary to consider other experimental tests of the modified kinematics (particle propagation studies, for example) which could disentangle a modification in the dispersion relation from a modification in the conservation law. 

The method we have presented can be used for any two body decay and it 
can also be extended to other processes. It is a first step of a systematic 
phenomenological analysis to establish bounds on (or identify) departures 
from SR in a framework going beyond EFT.

The main conclusion of the present work is that bounds extracted within the EFT scenario may be too naive (invalid) if the modified kinematics contains modified conservation laws.

\section*{Acknowledgments}
This work is supported by the Spanish MINECO grants FPA2012-35453 and CPAN-CSD2007-00042, and by DGIID-DGA grant 2013-E24/2. We acknowledge useful conversations with A. Grillo and F. M\'endez.

\appendix*

\section{Perfect colinearity at the threshold}

We show here that the threshold generated by the new kinematics in the UR limit for the two-particle decay, which is controlled by Eq.~\eqref{eq:UR-MKE}, is obtained at $\cos\theta=1$.
 
Considering Eq.~\eqref{eq:UR-MKE} as an implicit equation determining $(E_p+E_q)$ as a function of $(x,\cos\theta)$, we can take then the partial derivative of the equation with respect to $\cos\theta$ giving
\begin{multline}
\left[(1+x)(1-x)(1-\cos\theta)(E_p+E_q)-3\hat{\xi}_3^A(x)\frac{(E_p+E_q)^2}{\Lambda}\right]\frac{\partial(E_p+E_q)}{\cos\theta}\\ =\frac{(1+x)(1-x)}{2}(E_p+E_q)^2-\frac{1+x}{1-x}m_c^2-\frac{1-x}{1+x}m_d^2\,.
\label{eq:derivative}
\end{multline}
The right-hand side of Eq.~\eqref{eq:derivative} is positive, and the coefficient of the partial derivative in the left-hand side has the opposite sign to $\hat{\xi}^A$ when $\cos\theta=1$. This means that, in the case in which there is a threshold in a decay which is allowed in SR ($\hat{\xi}_3^A<0$ according to the comments of Section~\ref{UR}), $\partial{(E_p+E_q)}/{\partial{\cos\theta}}|_{\cos\theta=1}>0$, so that, for fixed $x$, $(E_p+E_q)$ takes its maximum value at $\cos\theta=1$. This maximum value of $(E_p+E_q)$ corresponds to an energy threshold above which the reaction ceases to be kinematically allowed. In the opposite case, when $\hat{\xi}_3^A>0$, we have a minimum of $(E_p+E_q)$ at $\cos\theta=1$, which corresponds to the threshold energy above which the reaction, forbidden in SR, is now allowed.

\end{document}